 \documentclass[preprint]{aastex}

\newcommand{\kms}{km s${}^{-1}$}
\newcommand{\ab}{$\sim$}

\newcommand{\p}{$\pm$}
\newcommand{\etal} {{et~al.}}

\newcommand{\HI} {H\kern0.1em{\sc i}}

\newcommand{\pks} {PKS\,2322$-$123}

\slugcomment{To appear in the Astrophysical Journal}

\begin{document}

\title{Redshifts and Neutral Hydrogen Observations of \\Compact
Symmetric Objects in the COINS Sample} 
\author{A. B. Peck\altaffilmark{1,2,3}, G. B. Taylor\altaffilmark{1},
C. D. Fassnacht\altaffilmark{1}, A. C. S. Readhead\altaffilmark{4} and
R. C. Vermeulen\altaffilmark{5}}
\altaffiltext{1} {National Radio Astronomy Observatory, P.O. Box O,
Socorro, NM 87801;\\apeck@nrao.edu,gtaylor@nrao.edu,cfassnac@nrao.edu}
\altaffiltext{2}{Physics Dept., New Mexico Institute of Mining and
Technology, Socorro, NM 87801}\altaffiltext{3} {Current Address: 
Max-Planck-Institut f\"{u}r Radioastronomie, D-53121 Bonn, Germany}
\altaffiltext{4}{California Institute of Technology, Dept. of
Astronomy, 105-24, Pasadena, CA, 91125}
\altaffiltext{5}{NFRA, Postbus 2, 7990AA Dwingeloo, The Netherlands}
\setcounter{footnote}{0}

\begin{abstract}

  Compact Symmetric Objects (CSOs) are young radio galaxies whose jet
  axes lie close to the plane of the sky, and whose appearance is
  therefore not dominated by relativistic beaming effects.  
  The small linear sizes of CSOs make them valuable for
  studies of both the evolution of radio galaxies and testing unified
  schemes for active galactic nuclei (AGN).  A parsec-scale region of
  gas surrounding the central engine is predicted by both accretion
  and obscuration scenarios.  Working surfaces, or ``hot spots,'' and
  the radio jets of CSOs are close enough to the central engines that
  this circumnuclear gas can be seen in absorption.

  The CSOs Observed in the Northern Sky (COINS) sample is comprised of
  52 CSO candidates identified in three VLBI surveys.  Of these, 27
  have now been confirmed as CSOs.  Optical redshifts are available in
  the literature for 28 of the CSO candidates, and \HI\ absorption has
  been detected toward four.  We present new optical spectroscopic
  redshifts for three of the candidates and summarize the current
  status of optical identifications.  We further report on the
  discovery of \HI\ in absorption towards the CSO J1816+3457 and
  summarize the results of neutral hydrogen absorption studies of the
  sources in this sample.

\end{abstract}

\clearpage

\section{Introduction}

Unified schemes for active galactic nuclei (AGN) seek to explain
several classes of objects with a smaller number of source types
viewed at different orientations (see the review by \citet{ant93}).  A
critical ingredient of all unified schemes is an obscuring disk or
torus which hides the nucleus when the source is viewed edge-on.  An
accretion disk is also required to feed the AGN and probably plays a
role in collimating the bipolar jets.  Although the composition of the
obscuring material in the torus may be varied, it seems likely that at
some radii and scale heights, there will be significant amounts of
neutral atomic hydrogen gas \citep{neu95}.  This should be detectable
in \HI\ absorption towards the core and inner jets of radio-loud
AGN. Sources with symmetric parsec-scale jets are ideal for testing
the model. Broad ($>$100 \kms) lines are expected this close to the
bottom of the potential well of the galaxy. Such broad absorption
lines have thus far been detected towards a few moderate power radio
galaxies such as Hydra~A \citep{tay96}, 3C\,84 \citep{cra82} and \pks\
\citep{tay99}, and in the compact symmetric objects (CSOs) 0108+388
\citep{car98}, 1146+586 \citep{gor89,pec98}, PKS 1413+135
\citep{car92} and 1946+708 \citep{pe298}.

In the inner regions of the disk a large fraction of the gas must be
ionized by the central engine.  This will result in free-free
absorption of the radio continuum at frequencies below $\sim$5 GHz as
seen in 3C\,84 \citep{wal98}, Hydra~A \citep{tay96}, 1946+708
\citep{pe298}, and \pks\ \citep{tay99}.  Another result of
the dense ionized gas, if it is magnetized, could be extremely high
Faraday rotation measures (RMs).  Owing to the lack of any polarized
flux in these systems, it has not been possible to directly measure
the RMs in any source with \HI\ absorption, but the presence of very
high RMs could explain why no polarized flux is detected from these
sources \citep{tay99, pec99}.

We are engaged in an ongoing project to use VLBI observations of \HI\
absorption to study the densities, kinematics, and scale heights of
the neutral gas in the obscuring torus in a moderate size sample of
compact objects.  This sample is known as the CSOs Observed in the
Northern Sky (COINS) sample, although to date only 27 have been
securely classified as CSOs.  Since CSOs are rare (\ab2\% of compact
objects; Peck \& Taylor 1999), it is necessary to start with large
VLBI surveys and go to moderately low flux density levels (\ab100 mJy
at 5 GHz) in order to obtain the 52 CSO candidates that make up the
COINS sample.  Follow-up multifrequency and polarimetric VLBI
observations must then be made in order to identify the center of
activity in each source and thus distinguish the true CSOs from core
jet sources with ``compact double'' morphologies.  Once the CSOs are
identified, their redshifts must be determined before they can be
searched for \HI\ in absorption.  Low spatial resolution observations
can then be carried out at the frequency of the redshifted neutral
hydrogen line, and detections can be followed up with extremely high
spatial and spectral resolution VLBI studies.  Multifrequency
continuum observations can also be used to image the free-free
absorption in suitable candidates, providing an additional means of
determining the geometry of the circumnuclear material.

We present the current status of the COINS survey project in $\S$2\ of
this paper.  $\S$3\ outlines our recent observations which provide
three new redshifts and one new \HI\ detection.  These new results,
and their importance to future work, are discussed in $\S\S$ 4 and 5.

\section{The COINS Sample}

The sources in the COINS sample have been identified based on images
in the Pearson-Readhead (PR; Pearson \& Readhead 1988),
Caltech-Jodrell Bank (CJ; Polatidis \etal\ 1995; Taylor \etal\ 1994)
and VLBA Calibrator (VCS; Peck \& Beasley 1997) Surveys.  The majority
of the CSO candidates were chosen from the VCS based on criteria
outlined in \citet{pec99}, while the sources chosen from the PR and CJ
surveys conformed to very similar criteria described in \citet{rea96}
and in \citet{trp96}.

The sources in the COINS sample are described in Table 1. Column (1)
lists the J2000 convention source name of the CSO candidate.  Column
(2) provides an alternate name, with those prefaced by PR or CJ
indicating selection from that survey.  Columns (3) and (4) show the
optical identification and magnitude of the source.  Column (5) lists
the redshift of the source, references for which are provided in the
table caption.  The last column in Table 1 lists the status of the
\HI\ absorption detections toward the source, providing the optical
depth or upper limit of any \HI\ absorption observation published to
date.

\section{Observations and Data Reduction}

The optical data were taken in two observing runs on the 200 inch
telescope at Palomar Observatory, 1998 March 30 and 1999 April 10.
Both runs used the Double Spectrograph with a 5200 \AA\ dichroic
beamsplitter and a 2\arcsec\ slit.  All target sources were observed
for 1500 seconds each.  The total effective wavelength coverage was
approximately 3500--9200 \AA\ with a resolution of 4.9 \AA.  The
spectra were extracted using standard IRAF techniques.
Wavelength calibration was performed using exposures of arc lamps
taken at intervals throughout the observations.  Observations of the
standard star Feige 34 were used to remove the response function of
the chip.  The conditions were not photometric on either run so the
flux calibration in the spectra shown should not be regarded as
absolute.

  On 1998 December 6, the source J1110+4817 was observed for 6 hours
on the Westerbork Synthesis Radio Telescope (WSRT) using the UHF-high
receivers in dual linear polarization. With the new DZB correlator we
obtained 256 spectral channels over a 10 MHz bandwidth (formal
resolution 47 kHz), centered at the frequency of the \HI\ line predicted
by the optical redshift. There were 10 operational telescopes.
Calibration of the bandpass shape and the flux density scale was based
on a brief scan of 3C286 (1328+307). The initial phase calibration was
refined by a few self-calibration and modelfitting loops, in which the
20 brightest continuum sources in the field were found. The spectrum
of J1110+4817 shown in Fig. 2a was then produced by vector averaging
(over time and baselines) all of the cross-correlation spectra;
J1110+4817 is unresolved with the WSRT and was at the phase center of
the data. The spectrum of J1816+3457 shown in Fig. 2b was obtained in
an analogous procedure from a 12-hour observation on 1999 June 17.

\section{Results}

\subsection{Redshifts} 
The spectra of the three CSO candidates for which redshifts were
obtained are shown in Figure 1.  Gaussian fits to the spectral lines
used to determine the redshifts for the three new CSO candidates are
summarized in Table 2.

{\bf J1111+1955} is a galaxy with a redshift of 0.299\p0.001.  Some
indication of stellar absorption features is seen (Ca H \& K, Balmer
break), but these fall at the location of the dichroic and so
the identifications are tentative.

{\bf J1311+1417} is a quasar at a redshift of 1.995\p0.003.

{\bf J1816+3457} is a radio galaxy at a redshift of 0.2448\p0.0003.  Here
again, stellar features are visible, but have not been fitted.

\subsection{Neutral Hydrogen}
Figure 2 shows the spectra of the redshifted 21cm hydrogen toward two
of the CSO candidates.

{\bf J1110+4817} has been identified as a quasar by \citet{hoo96}.  The upper
limit for \HI\ absorption at the redshift of the source is
$\tau\le$0.009. 

{\bf J1816+3457} exhibits \HI\ absorption with an optical depth of
$\tau$\ab0.035.  This line is centered at 1.1418 GHz (c$z$=73151
\kms).  Applying the relativistic correction required at this
redshift, the radial velocity of the \HI\ absorption line is
$v$=64434\p10 \kms, which is 184 \kms\ blueward of the optical
emission line radial velocity of $v$=64618\p 70 \kms.  This difference
of a couple of hundred \kms\ should not be overinterpreted, given that
the optical redshift is determined from emission lines, which can be
influenced by outflow or infall.  A VLBI study of the \HI\ absorption
in this source is currently underway.

\section{Summary}

The compact size, orientation, and high rate of \HI\ detection in
compact symmetric objects make these sources highly valuable in the
study of accretion, evolution and the unified scheme of AGN.
Unfortunately, the scarcity of these sources thus far has resulted in
too few extensive studies on which to base any general conclusions.
The COINS survey is an attempt to ameliorate this situation by
identifying a larger sample of CSOs which can be comprehensively
studied using VLBI techniques.  The first phase of this project, the
identification at radio wavelengths of CSO candidates from large, high
resolution surveys, has been completed.  Multi-frequency VLBI
follow-up observations to eliminate core-jet sources masquerading as
CSOs in the finding survey have been carried out by \citet{pec99}.

A requisite next step in the process is to obtain the redshift of the
COINS sources by optical spectroscopy.  Including the three redshifts
presented herein, our redshift completeness for the sample is 32 of 52
(61\%).  Of these 32, low spatial resolution (kpc-scale) observations
to look for the presence of \HI\ absorption are available for 6
sources (4 referred to in the introduction, 2 new ones presented
here). J1110+4817 is the only 1 of the 6 in which there is no
absorption line exceeding 1\% peak depth. Clearly, \HI\ studies are
very profitable for CSOs. The detection rate in this class of sources
is far higher than that found in the nearby radio galaxies of "normal"
size surveyed for \HI\ absorption by \citet{gor89}, which
yielded 4 detections for 29 galaxies.


Continuing work on this project involves using optical spectroscopy to
obtain redshifts for the remaining CSOs, many of which are extremely
faint at optical wavelengths.  Once this has been accomplished,  \HI\
absorption studies can be undertaken, providing a necessary complement
to free-free absorption and jet expansion studies. 

\begin{acknowledgements}

The National Radio Astronomy Observatory is a facility of the National
Science Foundation operated under a cooperative agreement by
Associated Universities, Inc.  AP is grateful for support from NRAO
through the pre-doctoral fellowship program.  AP also acknowledges the
New Mexico Space Grant Consortium for partial publication costs.  IRAF
is distributed by the National Optical Astronomy Observatories, which
are operated by the Association of Universities for Research in
Astronomy, Inc., under cooperative agreement with the NSF.  This
research has made use of the NASA/IPAC Extragalactic Database (NED)
which is operated by the Jet Propulsion Laboratory, California
Institute of Technology, under contract with the National Aeronautics
and Space Administration.

\end{acknowledgements}

\clearpage


\begin{figure}
\vspace{18.5cm}
\includegraphics{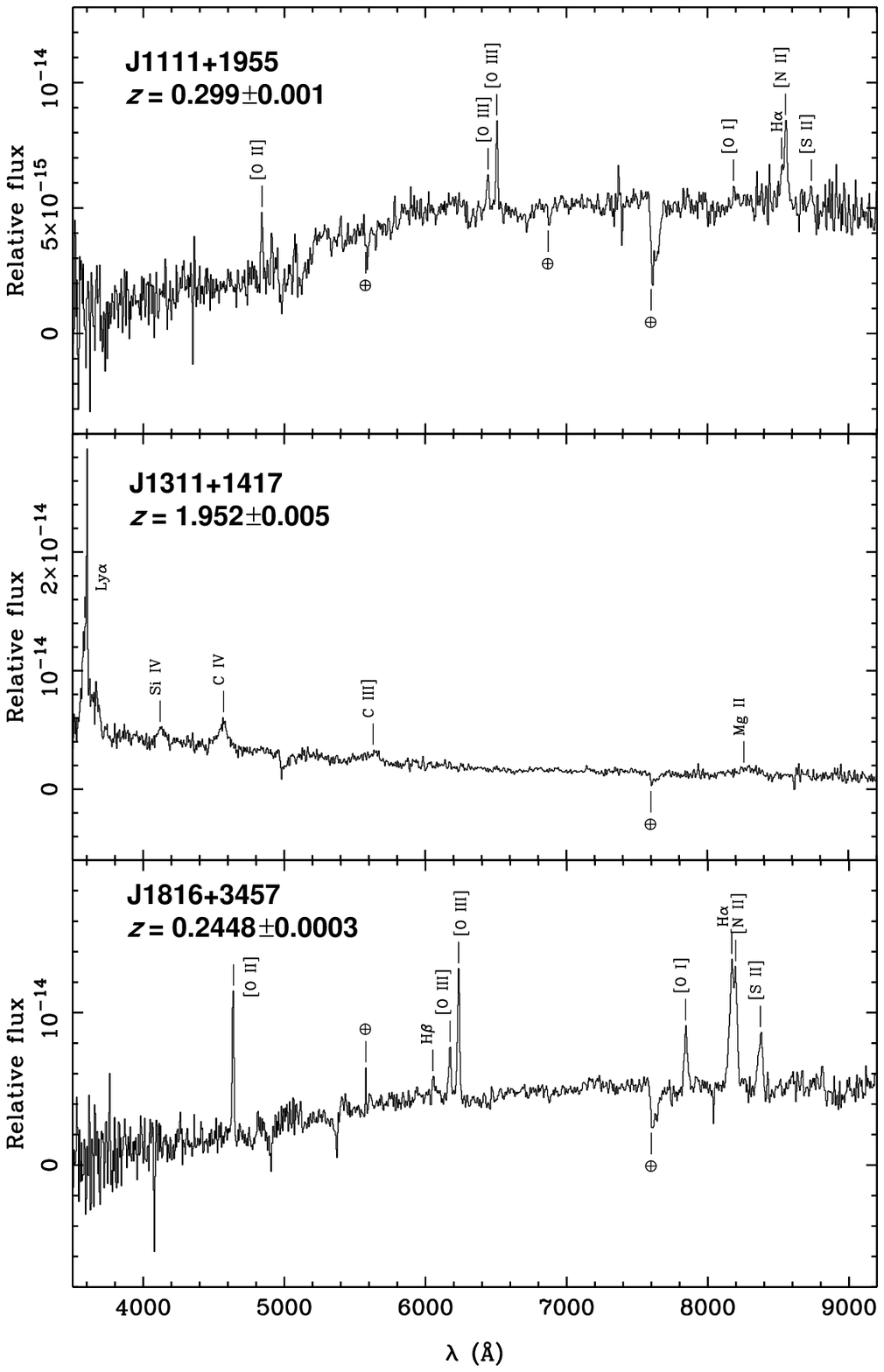}
\figcaption{Redshifts obtained for three of the CSO candidates.
Gaussian fits to the lines used in redshift determination are shown in
Table 2.
\label{fig1}}
\end{figure}

\begin{figure}
\vspace{16.5cm} \includegraphics{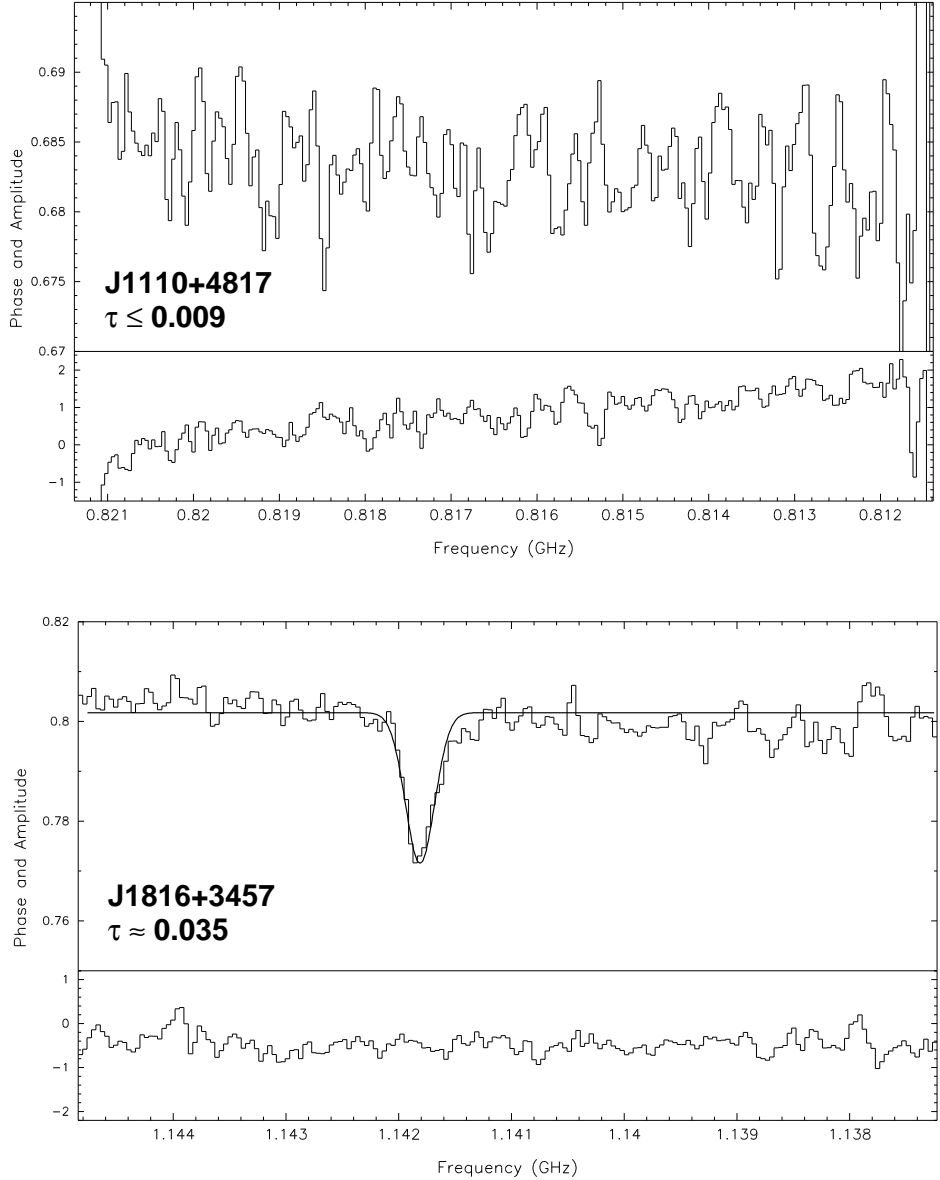} 
\figcaption{WSRT spectra obtained for two
of the CSO candidates. (a) The spectrum of J1110+4817 represents a six
hour observation.  The rms noise in each channel was \ab5 mJy/beam.
The upper limit on \HI\ optical depth is $\tau\le$0.009.  (b) The
spectrum of J1816+3457 represents a twelve hour observation.  The rms
noise in each channel was \ab5 mJy/beam.  The optical depth of the
\HI\ is $\tau$\ab0.035.  The line is centered on c$z$=73151 \kms~ and
has a FWHM of \ab80 \kms.
\label{fig2}}
\end{figure}

\begin{center}
\begin{deluxetable}{llcrrr}
\tabletypesize{\footnotesize}%
\tablecolumns{6}
\tablewidth{0pc} 
\tablecaption{The COINS Sample\label{tab1}}
\tablehead{
\colhead{Source Name} &\colhead{Alternate Name }&\colhead{ID}  &\colhead{$m$} &\colhead{$z$}  &\colhead{$\tau_{\rm HI}$} \\
\colhead{(1)} &\colhead{(2)} &\colhead{(3)} &\colhead{(4)}  &\colhead{(5)} &\colhead{(6)}}
\startdata

J0000+4054 &4C 40.52&G   &20.0&... &... \\
J0003+4807 &&...&EF  &... &... \\
J0029+3456&CJ 0026+346&G&20.4&... &... \\
J0029+0509& &QSO &20.0&1.63$^{[1]}$&... \\
J0111+3906&PR 0108+388&G&22.0&0.6703$^{[2]}$& 0.44$^{[3]}$ \\
J0132+5620&&... &EF  &... & ...   \\
J0204+0903&&... &EF  &... & ...  \\
J0207+3152&5C 06.008&G   &16.8&1.21$^{[4]}$& ... \\
J0332+6753&&... &EF  &... & ... \\
J0400+0550& &... &18.5&... &... \\
J0410+7656&PR 0404+768&G&22.0&0.5985$^{[2]}$&... \\
J0424+1442& &... &EF  &... & ...  \\
J0427+4133& &... &EF  &... & ... \\
J0518+4730& &... &EF  &... & ... \\
J0620+2102& &... &EF  &... & ... \\
J0650+6001&CJ 0646+600&QSO&18.9&0.455$^{[5]}$&... \\
J0713+4349&PR 0710+439&G&19.7&0.5180$^{[2]}$&... \\
J0753+4231&CJ 0749+426&QSO&18.1&3.59$^{[6]}$ &... \\
J0754+5324 &&... &EF  &... &  ... \\
J0842+1835 &&QSO &16.4&1.27$^{[7]}$&  ... \\
J0925+0019& &QSO &18.1&1.72$^{[7]}$& ... \\
J0954+7435 &&G   &21.7&... &... \\
J1035+5628&PR 1031+567&G&20.3&0.4597$^{[2]}$&... \\
J1107+7232& &QSO &18.9&2.10$^{[7]}$&... \\
J1110+4817&&QSO &19.2&0.74$^{[6]}$& $<$0.009$^{[8]}$ \\
J1111+1955& &G   &18.5&0.299$^{[8]}$ &... \\
J1143+1834& &... &EF  &... & ...  \\
J1148+5924&CJ 1146+596&G&11.0&0.011$^{[9]}$&0.04$^{[10]}$\\
J1227+3635& &QSO &21.5&1.97$^{[11]}$& ...  \\
J1244+4048&CJ 1242+410&QSO&19.0&0.813$^{[11]}$&... \\
J1311+1417& &QSO &19.5&1.952$^{[8]}$ & ...  \\
J1311+1658& &... &EF  &... & ...  \\
J1357+4353&CJ 1355+441&G&21.6&... & ...  \\
J1400+6210&PR 1358+624&G&20.9&0.4310$^{[2]}$&... \\
J1414+4554& &G   &19.9&0.19$^{[12]}$& ...  \\
J1415+1320&PKS1413+135&QSO &20  &0.25$^{[13]}$& 0.34$^{[14]}$ \\
J1546+0026& &G   &20  &0.55$^{[15]}$&... \\
J1651+0129& &... &21.2&... &  ...  \\
J1734+0926& &G   &20.7&... &   ...  \\
J1737+0621& &QSO &17.9&1.21$^{[9]}$&  ...   \\
J1810+5649&7C1809+5648&... &18.0&... &  ... \\
J1815+6127&CJ 1815+614&QSO&22.0&0.601$^{[16]}$&...  \\
J1816+3457& &G &18.7&0.245$^{[8]}$& 0.035 $^{[8]}$   \\
J1826+1831& &... &EF  &... & ...   \\
J1823+7938&CJ 1826+796&G&16.7&0.224$^{[17]}$ &... \\
J1845+3541&CJ 1843+356&G&21.9&0.764$^{[18]}$ &... \\
J1944+5448&CJ 1943+546&G&17.6&0.263$^{[19]}$&... \\
J1945+7055&CJ 1946+708&G&18.0&0.101$^{[19]}$& 0.05$^{[20]}$\\
J2022+6136&PR 2021+614&QSO&19.5&0.2280$^{[2]}$& ...\\
J2203+1007 &&... &EF  &... & ...  \\
J2245+0324 &&QSO &19.0&1.34$^{[21]}$& ... \\
J2355+4950&PR 2352+495&G&20.1&0.2383$^{[2]}$& ... \\
\tableline
\enddata

\tablecomments{The initial list of CSO candidates observed in the
northern sky: References are as follows: (1) Perlman \etal\ 1998; (2)
Lawrence \etal\ 1996; (3) Carilli \etal\ 1998; (4) Willott \etal\
1998; (5) Stickel \& K\"{u}hr 1993a; (6) Hook \etal\ 1996; (7) Hewitt,
A. \& Burbidge, G. 1988; (8) this paper; (9) Karachentsev 1980; (10)
van Gorkom \etal\ 1989; (11) Xu \etal\ 1994; (12) Falco \etal\ 1998;
(13) McHardy \etal\ 1991; (14) Carilli et al 1992; (15) Heckman \etal\
1994; (16) Vermeulen \& Taylor 1995; (17) Henstock \etal\ 1997; (18)
Vermeulen \etal\ 1996; (19) Stickel \& K\"{u}hr 1993b; (20) Peck
\etal\ 1999; (21) Wolter \etal\ 1997.}
\end{deluxetable}
\end{center}

\clearpage
\begin{center}
\begin{deluxetable}{lcrrrrr}
\tabletypesize{\footnotesize}%
\tablecolumns{7}
\tablewidth{0pc} 
\tablecaption{Redshifts for CSO candidates\label{tab2}}

\tablehead{
\colhead{Source Name} & \colhead{$z$} & \colhead{Integ.} & \colhead{Line ID} &
\colhead{$\lambda_{\rm obs}$} &\colhead{EW} &\colhead{FWHM} \\
\colhead{} & \colhead{(s)} &\colhead{}  & \colhead{}  &\colhead{(\AA)} &\colhead{(\AA)} & \colhead{(\AA)}}

\tablenum{2}
\startdata

J1111+1955&0.2991\p0.001&1500&[O II] $\lambda$3727.6& 4840&20.20 &15.6\\
&&& [O III] $\lambda$4960.3&6443&5.48&18.4 \\
&&& [O III] $\lambda$5008.2&6507&10.70&13.8 \\
&&& [O I]$ \lambda$6300.0&8189&3.50&20.6 \\
&&& H $\alpha$~$\lambda$6562.8&8531& Blend & \\
&&& [N II] $\lambda$6548.1&8556&Blend& \\
&&& [S II] $\lambda$6725.5&8733&4.5&25.4 \\
J1311+1417&1.952\p0.005&1500&Ly $\alpha$~$\lambda$1215.7&3596&102&40.5\\
&&&Si IV $\lambda$1396.7&4128&22.1&64.5 \\
&&&C IV $\lambda$1549.1&4568&43.1&70.8 \\
&&&C III] $\lambda$1908.7&5641&24.3&78.9 \\
&&&Mg II $\lambda$2799.8&8292&105&164.4 \\
J1816+3457&0.2448\p0.0003&1500&[O II]$\lambda$3727.6&4637 &78.7 &13.1 \\
&&& [O III] $\lambda$4960.3&6173&15.1&19.3 \\
&&& [O III] $\lambda$5008.2&6234&33.2&16.9 \\
&&& [O I]$ \lambda$6300.0&7844&26.9&30.1 \\
&&&H $\alpha$~$\lambda$6562.8&8172&Blend& \\
&&&[N II] $\lambda$6585.2&8200&Blend& \\
&&&[S II] $\lambda$6725.5&8371&44.9&44.8 \\

\tableline
\enddata
\end{deluxetable}
\end{center}


\begin{thebibliography}{}

\bibitem[Antonucci (1993)]{ant93} Antonucci, R. 1993, ARA\&A, 31, 473

\bibitem[Carilli, Perlman \& Stocke (1992)]{car92} Carilli, C. L.,
Perlman, E. S. \& Stocke, J. T. 1992, \apjl, 400, L13

\bibitem[Carilli \etal\ (1998)]{car98} Carilli, C. L., Menten, K. M.,
Reid, M. J., Rupen, M. P. \& Yun, M. S. 1998, \apj, 494, 175

\bibitem[Crane, van der Hulst \& Haschick (1982)]{cra82} Crane, P. C.,
van der Hulst, J. M., \& Haschick, A. D. 1982, in IAU Symp. 97,
Extragalactic Radio Sources, ed. D. S. Heeschen \& C. M. Wade
(Dordrecht:Reidel), 307

\bibitem[Falco, Kochanek \& Mu\~{n}oz (1998)]{fal98} Falco, E. E.,
Kochanek, C. S. \& Mu\~{n}oz, J. A. 1998, \apj, 494, 47

\bibitem[van Gorkom \etal\ (1989)]{gor89} van Gorkom, J. H., Knapp,
G. R. Ekers, R. D., Ekers, D. D., Laing, R. A. \& Polk, K. S. 1989,
\aj, 97, 708

\bibitem[Heckman \etal\ (1994)]{hec94} Heckman, T. M., O'Dea, C. P.,
Baum, S. A. \& Laurikainen, E. 1994, \apj, 428, 65

\bibitem[Henstock \etal\ (1997)]{hen97} Henstock, D. R., Browne,
I. W. A., Wilkinson, P. N. \& McMahon, R. G. 1997, \mnras, 290, 380

\bibitem[Hewitt \& Burbidge (1989)]{hew89} Hewitt, A. \& Burbidge,
G. 1989, \apjs, 69, 1

\bibitem[Hook \etal\ (1996)]{hoo96} Hook, I. M., McMahon, R. G.,
Irwin, M. J. \& Hazard, C. 1996, \mnras, 282, 1274

\bibitem[Karachentsev (1980)]{kar80} Karachentsev, I. D. 1980, \apjs, 44, 137

\bibitem[Lawrence \etal\ (1996)]{law96} Lawrence, C. R., Zucker,
J. R., Readhead, A. C. S., Unwin, S. C., Pearson, T. J. \& Xu, W.
1996, \apjs, 107, 541

\bibitem[McHardy \etal\ (1991)]{mch91} McHardy, I., Abraham, R.,
Crawford, C., Ulrich, M-H., Mock, P. \& Vanderspeck, R. 1991, \mnras,
249, 742

\bibitem[Neufeld \& Maloney (1995)]{neu95} Neufeld, D. A. \& Maloney,
P. R. 1995, ApJ, 447, L17

\bibitem[Patnaik \etal\ (1992)]{pat92} Patnaik, A. R, Browne,
I. W. A., Wilkinson, P. N. \& Wrobel, J. M. 1992. \mnras, 254, 655

\bibitem[Pearson \& Readhead (1988)]{pr88} Pearson, T. J., \&
Readhead, A. C. S. 1988, ApJ, 328, 114

\bibitem[Peck, \& Beasley (1997)]{pec97} Peck, A. B. \& Beasley,
A. J. 1998, in {\it IAU Colloquium 164: Radio Emission from Galactic
and Extragalactic Compact Sources} eds. J. A. Zensus, G. B. Taylor and
J. M. Wrobel (PASP: San Francisco), Vol.\ 144, p.\ 155


\bibitem[Peck \& Taylor (1998)]{pec98} Peck, A. B. \& Taylor,
G. B. 1998, \apjl, 502, L23

\bibitem[Peck, Taylor \& Conway (1999)]{pe298} Peck, A. B., Taylor, G. B. \& Conway, J. E. 1999,
\apj, 521, 103

\bibitem[Peck \& Taylor (1999)]{pec99} Peck, A. B. \& Taylor, G. B. {\it
submitted}

\bibitem[Perlman \etal\ (1998)]{per98} Perlman, E. S., Padovani, P.,
Giommi P., Sambruna, R., Jones, L. R., Tzioumis, A. \& Reynolds,
J. 1998, \aj, 115, 1253

\bibitem[Polatidis \etal\ (1995)]{pol95} Polatidis, A. G., Wilkinson,
P. N., Xu, W., Readhead, A. C. S., Pearson, T. J., Taylor, G. B., \&
Vermeulen, R. C.  1995, ApJS, 98, 1

\bibitem[Readhead \etal\ (1996)]{rea96} Readhead, A. C. S., Taylor,
G. B., Xu, W., Pearson, T. J., Wilkinson, P. N., \& Polatidis,
A. G. 1996, ApJ, 460, 612

\bibitem[Stickel \& K\"{u}hr (1993a)]{sti93} Stickel, M. \& K\"{u}hr, H. 1993a, A\&AS, 101, 521

\bibitem[Stickel \& K\"{u}hr(1993b)]{stb93} Stickel, M. \& K\"{u}hr,
H. 1993b, A\&AS, 100, 395

\bibitem[Taylor \etal\ (1994)]{tay94} Taylor, G. B., Vermeulen, R. C.,
Pearson, T. J., Readhead, A. C. S., Henstock, D. R., Browne, I. W. A.,
\& Wilkinson, P. N.  1994, ApJS, 95, 345

\bibitem[Taylor (1996)]{tay96} Taylor, G. B. 1996, ApJ, 470, 394

\bibitem[Taylor, Readhead \& Pearson (1996)]{trp96} Taylor, G. B.,
Readhead, A. C. S., \& Pearson, T. J.  1996, ApJ, 463, 95

\bibitem[Taylor \etal\ (1999)]{tay99}Taylor, G. B., O'Dea, C. P.,
Peck, A. B. \& Koekemoer, A. M. 1999, ApJ, 512, L27


\bibitem[Vermeulen \& Taylor (1995)]{ver95} Vermeulen, R. C. \& Taylor, G. B. 1995, \aj, 109, 1983

\bibitem[Vermeulen \etal\ (1996)]{ver96} Vermeulen, R. C., Taylor,
G. B., Readhead, A. C. S. \& Browne, I. W. A. 1996, \aj, 111, 1013

\bibitem[Walker \etal\ (1998)]{wal98} Walker, R.~C., Kellermann,
K.~I., Dhawan, V., Romney, J.~D., Benson, J.~M., Vermeulen, R.~C., \&
Alef, W. 1998, in {\it IAU Colloquium 164: Radio Emission from
Galactic and Extragalactic Compact Sources} eds. J. A. Zensus,
G. B. Taylor and J. M. Wrobel (PASP: San Francisco), Vol.\ 144, p.\
133

\bibitem[Willott \etal\ (1998)]{wil98} Willott, C. J., Rawlings, S.,
Blundell, K. M. \& Lacy, M 1998, \mnras, 300, 625

\bibitem[Wolter \etal\ (1997)]{wol97} Wolter, A., Ciliegi, P., Della
Ceca, R., Gioia, I. M., Henry, P., Maccacaro, T., Padovani, P. \&
Ruscica, C. 1997, \mnras, 284, 225

\bibitem[Xu \etal\ (1994)]{xuw94} Xu, W., Lawrence, C. R., Readhead,
A. C. S. \& Pearson, T. J. 1994, \aj, 108, 395

\end{thebibliography}
\end{document}